\documentclass[12pt,preprint]{aastex}
\shorttitle{The Old Nova V603 Aql} 
\shortauthors{Sion et al.} 

\begin{document}

\title{Far Ultraviolet Spectroscopy of Old Novae I. 
V603 Aquila\altaffilmark{1}}

\author{Edward M. Sion, Patrick Godon\altaffilmark{2}, Alexandra Bisol} 
\affil{Astrophysics \& Planetary Science, Villanova University, \\
800 Lancaster Avenue, Villanova, PA 19085, USA}
\email{edward.sion@villanova.edu, patrick.godon@villanova.edu, 
alexandra.bisol@villanova.edu} 

\altaffiltext{1}{Based on observations made with the NASA-CNES-CSA 
{\it Far Ultraviolet Spectroscopic Explorer (FUSE)}. {\it FUSE} was operated 
for NASA by the Johns Hopkins University under NASA contract 
NAS5-32985. }  

\altaffiltext{2}{Visiting at the Henry Rowland Department of Physics,
The Johns Hopkins University, Baltimore, MD.} 

\begin{abstract} 
We present the results of a synthetic spectral analysis of the far 
ultraviolet archival IUE, HST and FUSE observations of the fast old nova 
V603 Aql,  obtained some 90 years after its 1918 nova outburst. 
Our analysis utilizes the new Hubble 
FGS parallax distance for this nearly face-on old nova, a high white 
dwarf mass and a low reddening. Our analysis includes non-truncated 
optically thick accretion disks since V603 Aql is neither a polar nor an 
intermediate polar. Our synthetic spectral modeling of the FUSE 
and HST spectra analyzed separately 
indicate a mass transfer rate 
$\dot{M}  = 1.5-2.2\times 10^{-9}M_{\odot}$/yr  
for the FUSE and HST spectra respectively, assuming a WD mass
of $1.2M_{\odot}$. The mass accretion rate  
also depends on the assumed WD mass, and increases by a factor of 
two for a WD mass of $0.8M_{\odot}$. Combining the FUSE and HST spectra
together lead to the same results.  
Potential implications are discussed.   
\end{abstract}


\section{Introduction}

There is much that we do not know about old novae (post-classical 
novae in, or approaching, quiescence). Do they become dwarf novae as their 
accretion rates drop? When does nuclear burning and hence the soft 
X-ray production stop following the nova explosion? What are the 
accretion rates of old novae as a function of time since the nova 
explosion? Is there enhanced mass transfer due to irradiation of 
the secondary donor star (causing the donor to bloat and/or by 
driving a wind off of the donor star) by the hot white dwarf and/or 
hot accretion disk? 

All of the above questions tie into a critical test of whether or not
the hibernation theory of cataclysmic variable evolution is viable 
(Shara et al. 1986). Hibernation theory posits that the mass
transfer in CVs declines after the nova occurs because the secondary 
star detaches from its Roche lobe due to the angular momentum loss from the nova explosion. These systems would become unobservable (extinct)
until magnetic stellar wind mass loss and gravitational wave radiation
lead to the resumption of Roche lobe overfill by the donor star, 
and mass transfer. A knowledge of the mass transfer behavior of 
old novae as a function of time since the outburst is a potentially 
critical test of hibernation theory. Thus far, this theory has not 
been supported by long term optical observations (e.g. \citet{sch14a,sch14b}). The results of testing hibernation theory up to now have been mixed.

Soon after the nova outburst, the post-nova will 
radiate at a very high bolometric luminosity (but not quite the 
Eddington limit) until hydrogen is destroyed by nuclear reactions, 
thus causing the H shell burning to stop. The time scale for this 
to occur is predicted to be in the range of 1 year  to $>$ 10 years, 
possibly as long as 300 years depending upon the mass of the WD and 
the amount of H left on the WD after the nova explosion, and is 
inversely dependent on the white dwarf mass. As mass is lost and 
the envelope shrinks, the effective temperature increases until 
exceeding $10^5$ K, and the peak emission moves to the UV and then 
to the EUV and soft X-ray ranges. Therefore, in these spectral 
regions the nova turn-off should be best observed \citep{sel89}.

The old nova V603 Aql (Nova Aquilae 1918) offers a number of advantages for a synthetic 
spectral analysis. This is important because it allows us to constrain 
the accretion rate in old novae by modeling the accretion disk with 
synthetic disk spectra. The accretion disk light is typically the 
dominant source of FUV flux in these systems.

Fortunately, V603 Aql has an accurate trigonometric parallax and far 
ultraviolet archival HST and FUSE observations.
Our analysis utilizes the new Hubble FGS 
parallax distance of 249 pc $^{+9}_{-8}$pc \citep{har13} 
which removes one critical free parameter in model fitting. The system has had 
many fairly good quality far ultraviolet spectra obtained with IUE, an HST 
STIS spectrum \citep{sel13} covering much of the far UV wavelength range covered with 
IUE and also a FUSE spectrum \citep{sel13} enabling the model fitting to extend down to the 
Lyman Limit. It is known to have a low orbital inclination ($ i = 13-15^{\circ}$) 
making it nearly face-on to the observer \citep{rit98}, a range of possible white dwarf mass from $0.9 M_{\odot}$ (Warner 1976) to
$1.2 \pm 0.2 M_{\odot}$, raising the possibility that it could eventually 
be a SN Type Ia progenitor, and it has a low reddening E(B-V) = 0.07-0.10
(\citet{gal74}; this paper). The physical and orbital parameters from the literature 
are summarized in Table 1. In view of these observed and derived parameters, 
we seek to determine or tightly constrain the accretion rate of V603 Aql and any other
properties that we can derive from synthetic spectral modeling of the system. 

\section{Far Ultraviolet Archival Spectra}

There are 110 far ultraviolet spectra obtained for V603 Aql with the short 
wavelength prime camera (SWP) on the IUE spacecraft. 
The IUE SWP spectra cover the wavelength range 1150-1978\AA\ , 
while the IUE LWP spectra cover the 1885-3165\AA\ spectral range.  
However, because 
V603 Aql has very good and high S/N HST/GHRS archival spectra, 
we do not use the IUE spectra for the spectral fits. Instead we
use IUE SWP+LWP spectra to assess the reddening E(B-V) 
using the method of \citet{ver87}. 
Namely, we deredden a spectrum covering the 2175\AA\ 
`bump` for different values of E(B-V) and then we inspect the resulting
dereddened spectra. The dereddened spectrum for which the  2175\AA\  
`bump` vanishes indicates the E(B-V) value. In Fig. 1 we show the
actual dereddened spectra. We find a reddening 
E(B-V)=$0.10$ which we adopt in this work.  
Using the 2175\AA\  
feature to estimate E(B-V) introduces an error of up to 
20\% \citet{fit99}. 
Consequently we have $E(B-V)=0.1 \pm 0.02$.  
All of the HST and FUSE fluxes presented here were de-reddened 
using the IUERDAF IDL routine, UNRED, with $E(B-V) = 0.10$.
In the discussion we evaluate the error on $\dot{M}$ due to an 
error on $E(B-V)$. 

V603 Aql was observed with FUSE over seven FUSE orbits in June 2002.
Since the inclination is low, we decided to combine the exposures
together without considering possible changes due to 
the orbital phase. Nonetheless, we checked the
individual exposures (FUSE orbits) of the spectrum and found that 
the continuum flux
level varies by no more than $\sim $10\% from one orbit to the other.  
The largest variation
occur near the O\,{\sc vi} doublet and Ly$\beta$ region, possibly due to 
the combined effects of broad absorption and emission of these species. 
The change only mildly affects the absorption lines. 
We note, in passing, 
that IUE observations have shown \citet{bor03} that both the continuum
flux level and the emission lines (e.g. C{\sc iv} 1550) fluctuate 
with time and the `scatter` is of the order of 100\% in the 1992 IUE data
(note that it is only $\sim 30$\% in the 1989 IUE data).  
The FUSE observations were obtained on 2002-06-07 
with the LWRS in TIMETAG mode with a total exposure time of 16,807 s. 
FUSE has a spectral range covering the higher order of the Lyman series,
namely from $\sim 905$\AA\ 
to $\sim 1185$\AA\ .  
We combined these FUSE exposures together and then extracted the co-added 
spectrum from the combined fits files. 
We follow the procedure described in \citet{god12} to process the
FUSE spectra. Because the SiC channels (1aSiC, 1bSiC, 2aSiC, 2bSiC)
did not collect much data, the spectrum starts at 980\AA\ (instead of
$\sim$910\AA\ ) and it has a gap around 1085\AA\ .
The resulting combined FUSE 
spectrum is displayed in Fig.2 where the strongest absorption and 
emission  features are identified. Some significant interstellar 
absorption (molecular hydrogen) affects the spectrum.           

The HST GHRS spectrum of V603 Aql we use here is the combination of 
the z37v0204t and z37v0205t exposures which were obtained on 
1996-10-06 with the G140L/2.0 configuration.  
The first exposure covers the 1140-1435\AA\ 
spectral range, while the second exposure covers the 1367-1663\AA\  
spectral range.  
These pipeline-processed spectra were  
downloaded (as ascii/vo tables) from the   
virtual observatory using VOSpec.  
In Fig.3, we display the HST/GHRS combined z37v0204t+z37v0205t 
spectrum as flux $F_{\lambda}$ versus 
wavelength $\lambda$ in Angstroms - \AA\ , 
covering together the spectral region from 1140\AA\ 
to 1663\AA\ . 
Note the Ly$\alpha$ absorption line (which 
may have an interstellar contribution), steeply rising continuum 
and strong emission lines due to  C\,{\sc iii} (1175), 
Si\,{\sc iv} (1394, 1402), C\,{\sc iv} (1548, 1551) and He\,{\sc ii} (1640). 

\section{Synthetic Spectral Analysis}

We adopted model accretion disks from the optically thick, steady 
state disk model grid of \citet{wad98}. 
In these accretion disk models, 
the outermost disk radius, $R_{out}$, is chosen so that $T_{eff}(R_{out})$  
is near 10,000K since disk annuli beyond this point, which are cooler 
zones with larger radii, would provide only a very small contribution 
to the mid and far UV disk flux.
For the disk models, every combination of $\dot{M}$, inclination $i$ and 
white dwarf mass was fitted to the data. We selected those models 
with inclination angle $ i = 41, 18,~~ M_{wd} = 0.80, 1.03,  1.21$  
$M_{\sun}$ and $\log (\dot{M}(M_{\sun})$/yr) = -8.0, -8.5, -9.0, 
-9.5, -10.0, -10.5. For the WD models, we used TLUSTY Version 203 
\citep{hub88} and Synspec48 \citep{hub95}  to construct a 
grid of temperatures from 12,000K to 60,000K in steps of 1,000K to 5,000K,  
with $Log(g)$ corresponding to the white dwarf mass of the 
accretion disk model. 

The analysis of the FUV spectra of V603 Aql is strengthened by the 
fact that the distance is accurately  known (249 +8/-9 pc) thus removing 
one free parameter. The inclination is very low (13 - 15 degrees) 
and the white dwarf mass (0.9 to 1.2 $M_{\odot}$) is known to within 
$\pm 0.2 M_{\odot}$.  Since the distance is known, 
for a given WD mass (and therefore radius), 
the best-fitting accretion disk model is 
obtained simply by scaling the model to the distance  
published from Hubble FGS 
trigonometric parallax, namely 249 pc +8/-9 pc. Since we feel 
secure with both the distance and the low inclination of V603 Aql, 
essentially two critical free parameters are tightly constrained. 
We present the results of our synthetic spectral fitting with disks 
and photospheres in Section 4 where we have modeled the FUSE spectrum 
alone, the HST  spectrum alone and finally, the combination of the 
FUSE + HST spectrum to attempt to consistently
fit a broader wavelength baseline, than HST and FUSE individually.

\section{Synthetic Spectral Fitting Results}

We started by fitting the FUSE spectrum alone  
for the combination of parameters in our disk and white dwarf
model grids (see previous section) with the distance fixed at 249pc, 
and for white dwarf masses of 0.8, 1.0 and 1.2 $M_{\odot}$. 
A single WD model, without a disk, could not fit the data. 
Namely, the white dwarf is relatively massive and has a small
radius, and consequently its contribution to the overall flux was 
of the order of $\sim 1$\% and did not affect the results. 
Realistic best fits were obtained for disk models, and we     
included a moderately hot WD ($T_{wd}=30,000$K) for the sake of
completeness even though the WD did not affect the results.    
The mass accretion rate $\dot{M}$ of the disk model fit depended mainly
on the WD mass and we list $\dot{M}$ for each different value of $M_{wd}$ 
($0.8,~1.0,~1.2M_{\odot}$) in Table 3. We present the model fit
for the $M_{wd}=1 M_{\odot}$ case in Figure 4. The model has 
$\dot{M}= 2.4 \times 10^{-9} M_{\odot}$/yr, $i=18$deg (the lowest
inclination in our disk model grid) and $d=249$pc  The model fit, as explained
in the Figure, was carried out between the continuum unaffected by 
ISM absorption (in red) and the synthetic spectrum (solid black line).   
Excess flux appears in all the fitting and indicates broad emission
from N\,{\sc iii} (990), H\,{\sc i} (1026), O\,{\sc vi} (1032) 
C\,{\sc iii} (1175), and possibly from S\,{\sc iv} (1173). The 
C\,{\sc iii} and O\,{\sc vi} emission were masked before the fitting
as they were readily apparent while the other lines were not.  

Next we fit the HST/GHRS spectrum alone, masking the emission lines
and the bottom of the Ly$\alpha$ region. 
The model fit resulted in an accretion rate very similar to what we obtained 
by modeling the FUSE spectrum, as shown in Table 3 for the different
WD masses assumed. 
The model fit to the HRS spectrum is shown in Figure 5 for $M_{wd}=1M_{\odot}$. 
The model fit has the following parameters: $M_{wd}=1.0 M_{\odot}$, $i = 18$deg, and a corresponding accretion rate of $\dot{M}=3 \times 10^{-9} M_{\odot}$/yr. 
The overall slope of the observed spectrum is more shallow than the
slope of the synthetic spectrum, and this is apparent especially in the longer
wavelengths. 
This might be due to the contribution of a colder component
peaking in the NUV or optical which we are not modeling (see discussion),  
due to either the irradiated donor star or the hot spot at the 
outer rim of the accretion disk where the gas stream from the inner 
Lagrangian point impacts supersonically onto the disk.

Lastly, we fit the FUSE+HRS combined spectrum, to increase the wavelength
range. For this purpose we had to scale the spectra to each other
and multiplied the GHRS spectrum by a factor of 0.80.  
As before, we fixed the distance to 249 pc and obtained results
almost identical to the FUSE spectral fit results. 
In Table 3, we summarize the accretion rates derived from the fitting. 
In Figure 6, we display the fit for the $M_{wd}=1M_{\odot}$ 
model. For clarity we have intentionally removed the ISM molecular 
hydrogen modeling in the shorter wavelengths, but we kept it in 
the Ly$\alpha$ region. Here too we see that in the longer wavelengths
the observed spectrum has extra flux, indicating the possible presence
of a colder component.  

Overall however, we are satisfied that we have achieved a robust
value of the accretion rate of V603 Aql.  

For our accretion 
rate, and with the assumption that compressional heating alone is 
operating, we transform our rate of accretion to a white dwarf 
effective temperature. This yields $T_{eff} = 30,000$K. 
The accreting white dwarf in the system contributes only about 
$\sim 1$\% of the FUV flux.

\section{Discussion}

It is not surprising that the FUV spectrum of the old Nova V603 Aql 
is dominated by accretion light from its optically thick accretion 
disk since old novae generally appear to sustain high accretion 
rates due most likely to the heating and irradiation effects on 
the secondary star by the nova outburst. The FUV spectra we used 
were obtained in 1996 and 2002, 78 and 84 years after the 1918 
nova explosion. It is therefore not unusual to have an accretion 
rate as high as we have derived. \citet{pue07} found an accretion rate 
of $1.4 \times 10^{-9} M_{\odot}$/yr, using a different 
model-fitting method (statistical optimization) than ours.

Since our model fitting depends sensitively on the observed 
continuum  level and slope and there is an error range of reddening values E(B-V), 
namely $0.10 \pm 0.02$, it is of interest to explore how 
the corrected fluxes are affected, and hence the derived accretion rate. 
The range of error in E(B-V) (0.02) affects the corrected flux by a factor 
of 0.85, such that for E(B-V)=0.08, the flux is 
 decreased to 0.85 
the value it has for E(B-V)=0.10;
and for E(B-V)=0.12 the flux is increased to 1/0.85 (=1.17). 
To a first order estimate, the mass accretion itself is directly 
proportional to the flux, such that the error on $\dot{M}$, due 
to the reddening error, is $-0.15\dot{M}/+0.17\dot{M}$ (-15\% and +17\%).
 
As stated in Section 2, we used the IUERDAF script `unred' 
(IDL routine) which by default  uses the reddening law of Savage 
\& Mathis (1979) assuming R=3.1. This option gives the same results 
as the reddening law by Seaton (1979) and uses the same value of R, 
3.1, which is an average of the Galactic (Milky Way) reddening. 
Ideally the value of R has to be known in the direction of the object, 
as it varies in the Galaxy by more than a factor of 2 from about 
2.2 to 5.5 (Fitzpatrick 1999). The reddening itself,
E(B-V), varies like 1/R, and consequently the possible error in R introduces an additional error
of +41\% (=3.1/2.2 - 1.0) and -44\% (3.1/5.5 - 1.0) in the value of the reddening and in its law
(this is assuming that R could be as low as 2.1 or as high as 5.5).
We cannot rule out that the discrepancy between the observed spectrum and the model is due to
the reddening law we are using. Since R is unknown, however, it is standard practice to use the 3.1
value and the `standard' reddening law by \citet{sea79}.

Recent work by \citet{joh14} has shown that 
V603 Aql returned to deep quiescence by 1938 and is fading in the 
optical at a rate of 0.44 +/- 0.02 magnitudes per century. The 
Hibernation model of \citet{sha86} predicts that old novae should fade by 
roughly one magnitude per century. The gradual cooling of the white 
dwarf reduces the irradiation thus allowing the donor star to relax 
into thermal equilibrium again, eventually detaching from its Roche 
lobe and hence entering hibernation. There is little doubt that in V603 
Aql, there is enhanced mass transfer due to irradiation of the 
secondary donor star (causing the donor to bloat and/or by driving 
a wind off of the donor star) by the hot white dwarf and/or hot 
accretion disk. 
Since the FUSE spectrum was taken 6 years after the HST spectrum,
it is tempting to speculate the difference in mass accretion rate
obtained from fitting the FUSE and HST spectra is due to an 
actual decrease of the mass accretion rate (about 8\%). 
However, based on the IUE data, it is very likely that the change
in the continuum flux level between the HST data and the FUSE data
is solely due to the fluctuations of the UV source as described
by \citet{bor03}. 

Based upon the present data, we do not know 
how close V603 Aql is to the termination of its 
irradiation-induced, enhanced mass transfer 
(cf. \citet{tap13}). 
Since V603 Aql has been shown to be essentially 
non-magnetic \citep{bor03,muk05}, 
then the strong persistent He II (1640) emission line may originate 
in the hot inner disk region still irradiated by the heated central white dwarf.

V603 Aql was observed with ASCA,  
and a mass accretion rate of {\it only} $1.6 \times 10^{-10}M_{\odot}$/yr 
was derived by \citet{muk05}. This is much smaller than the value
we derived here, however,  
most CVs at high mass accretion rates
usually exhibit an X-ray derived mass accretion rate much smaller than derived 
from UV spectral analysis.  
A possible explanation is that the boundary layer is optically thin 
\citep{pop95} and  
cannot radiate efficiently. Its energy is advected into the outer
layer of the accreting white dwarf, thereby increasing its temperature.  
The HST-GHRS observations of \citet{fri97}                 
reveal the presence of a chromosphere-corona which surrounds the 
accretion disk and co-rotates with it which they associate with the emission lines
which are rotationally broadened. The blue-shifted absorption is the 
result of the wind outflow photo-ionized by the hot innermost disk/boundary layer. 
The inference of a disk corona in the above study 
suggests that the X-rays may arise from the corona.

\section{acknowledgements}

This work is supported by NASA grants NNX13AF12G and 
NNX13AF11G to Villanova University. We are grateful to an anonymous 
referee whose helpful comments have improved our paper. PG is thankful to William P. Blair
for his kind hospitality in the Henry A. Rowland 
Department of Physics and Astronomy 
at the Johns Hopkins University, Baltimore, MD.

\clearpage

\begin{deluxetable}{ccl} 
\tablewidth{0pt} 
\tablecaption{Physical and Orbital parameters of V603 Aql} 
\tablehead{ 
Parameter      &    Value      & Reference     \\ 
} 
\startdata 
Explosion Date     &  1918                &                \\  
Nova Speed Class   &  Fast                &                \\  
Orbital Period     & 3.317 h (0.138201 d) & \citet{rit98}     \\ 
Inclination        & 13-15 degrees        & \citet{rit98}     \\ 
Distance           & 330pc                & \citet{due87}       \\ 
                   & 360pc                & \citet{hub27}     \\ 
                   & 380pc                & \citet{mcl60}       \\ 
                   & 430pc                & \citet{har88}       \\ 
                   & 237pc                & Hipparcos Parallax   \\ 
                   & 249pc                & Hubble FGS parallax \citep{har13} \\
Visual Magnitude   &  11.64               & \citet{rit98}             \\ 
E(B-V)             &   0.07               & \citet{gal74}       \\ 
White Dwarf Mass   & 1.2$\pm 0.2M_{\odot}$& \citet{rit98}             \\ 
                   & 0.9 $M_{\odot}$      & Warner 1976)      \\ 
Donor star mass    & 0.29 $M_{\odot}$     & \citet{rit98}              
\enddata
\end{deluxetable}

\clearpage

\begin{deluxetable}{ccccccc} 
\tablewidth{0pt} 
\tablecaption{Archival Observations Log}                    
\tablehead{ 
Telescope/    & Filter/Grating & Obs.ID      & Obs.Date   & Obs.Time    &  Exp.time   & $\lambda$        \\ 
Instrument    &                              & DD/MM/YYYY & HH:MM:SS    &  $<$sec$>$  & $<$\AA\ $>$      \\
} 
\startdata 
HST/GHRS      & G140L/2.0     & z37v0204t   & 06/10/1996  & 07:25:34    &  1088       & 1367-1663         \\ 
HST/GHRS      & G140L/2.0     & z37v0205t   & 06/10/1996  & 07:49:02    &  544        & 1140-1435         \\ 
FUSE          & LWRS          & q1130101000 & 07/06/2002  & 16:27:19    &  16807      & 980-1188          \\ 
\enddata
\end{deluxetable}

\clearpage 

\begin{deluxetable}{cccc} 
\tablewidth{0pt} 
\tablecaption{Synthetic Spectral Fitting Results} 
\tablehead{ 
              & FUSE              &  HRS              &  FUSE+HRS                 \\ 
$M_{wd}$      &  $\dot{M}$        &  $\dot{M}$        & $\dot{M}$                \\
$<M_{\odot}>$ &  $<M_{\odot}/yr>$ &  $<M_{\odot}/yr>$ & $<M_{\odot}/yr>$         \\  
} 
\startdata 
 0.8 &  $3.2\times 10^{-9}$ & $4.0\times 10^{-9}$ &  $3.3 \times 10^{-9}$   \\  
 1.0 &  $2.4\times 10^{-9}$ & $3.0\times 10^{-9}$ &  $2.6 \times 10^{-9}$   \\ 
 1.2 &  $1.5\times 10^{-9}$ & $2.2\times 10^{-9}$ &  $1.5 \times 10^{-9}$    
\enddata
\end{deluxetable}

\clearpage

\begin{figure}
\vspace{-6.cm} 
\plotone{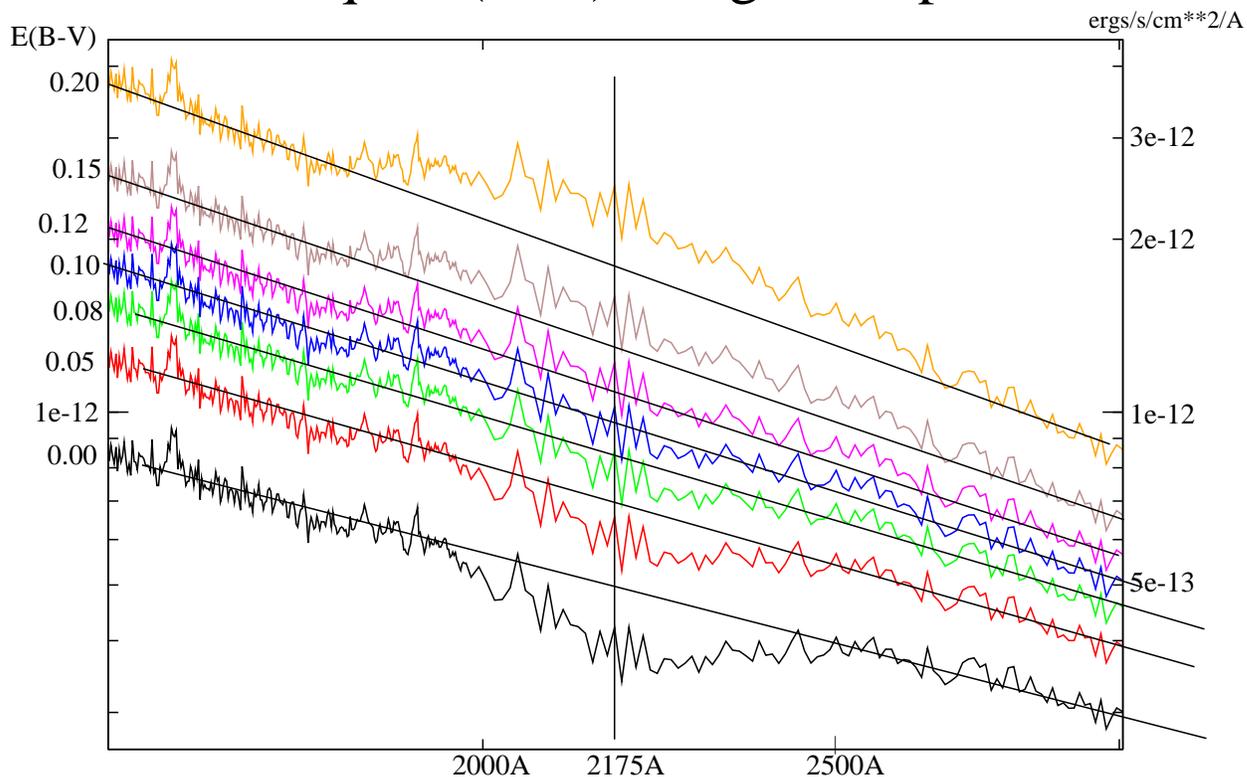} 
\caption{ 
The IUE spectrum of V603 Aql has been dereddened assuming different values
of E(B-V) as indicated on the left. A vertical line has been placed at
the location of the 2175\AA\ 
spectral feature  due to interstellar reddening. 
The bottom graph (solid black line) 
shows the observed spectrum without dereddening,
the spectral feature clearly shows as a broad absorption about 600\AA\ 
wide. The dereddened spectra are shown in colors for clarity.  
As the E(B-V) value is increased to 0.10, the 2175\AA\ 
spectral features disappears in the dereddened spectrum. 
As the E(B-V) value is increased further, the feature starts to show 
as a broad emission. 
Straight (solid black) lines have been drawn for easy comparison. 
From this graph we adopt $E(B-V)=0.10 \pm 0.02$.  
Note the spectrum shown here was generated by combining together several IUE
exposures with the same continuum flux level. 
}
\end{figure}

\begin{figure}
\vspace{-6.cm} 
\plotone{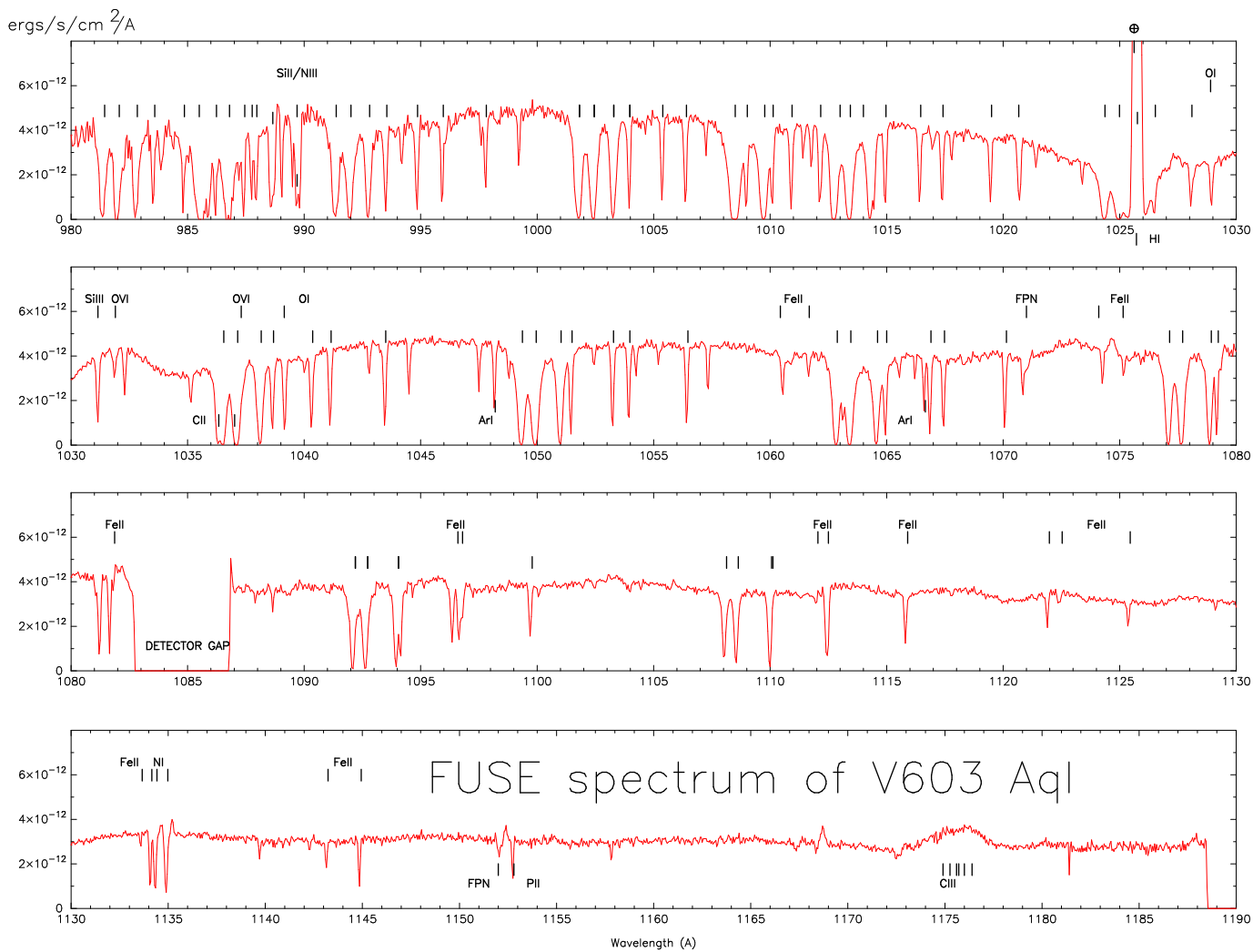} 
\caption{ 
The FUSE spectrum of V603 Aql has been dereddened  
assuming E(B-V)=0.10 and is shown with line identifications. 
The spectrum is strongly affected by sharp absorption lines.   
The prominent molecular hydrogen lines are identified with vertical
tick marks without label at 
$5 \times 10^{-12}$ergs$~$s$^{-1}$cm$^{-2}$\AA$^{-1}$.
Some low ionization species, such as e.g. Fe\,{\sc ii}, 
are also identified.   
The sharp absorption lines could be due to ISM absorption, 
or to the surrounding material ejected during previous eruptions,
or to a combination of both. 
The Si\,{\sc iii} and O\,{\sc vi} lines around 1030-1032\AA\ 
are probably from V603 Aql itself, as is  
the C\,{\sc iii} (1175) multiplet which appears as a
broad and shallow emission feature. 
Known FUSE fixed pattern noises are marked  'FPN'. Because of the failure of
one of the FUSE channels there is detector gap in the spectrum around 1085\AA . 
The sharp H\,{\sc i} (1025) emission lines are due to airglow.   
}
\end{figure}

\begin{figure} 
\plotone{f3.ps}   
\caption{ 
The HST GHRS G140L spectrum of V603 Aql has been dereddened
assuming E(B-V)=0.10.  
We identify all the lines as marked.
The N\,{\sc v} doublet ($\sim$1240\AA\ ) has been marked at its expected 
(rest frame) position but is not detected.  
The C\,{\sc iii} (1175), Si\,{\sc iv} (1400), C\,{\sc iv} (1550),  
He\,{\sc ii} (1640), and possibly C\,{\sc ii} (1335) 
lines are all in emission. 
Note the double absorption in the bottom of the
Ly$\alpha$, with a shift of $\pm 3$\AA\ 
corresponding to a projected velocity of $\sim \pm 750$km/s, 
or a rotational velocity of 2900km/s for $i=14$deg and 3330km/s
for $i=13$deg. 
} 
\end{figure} 

\begin{figure} 
\vspace{-5.cm} 
\plotone{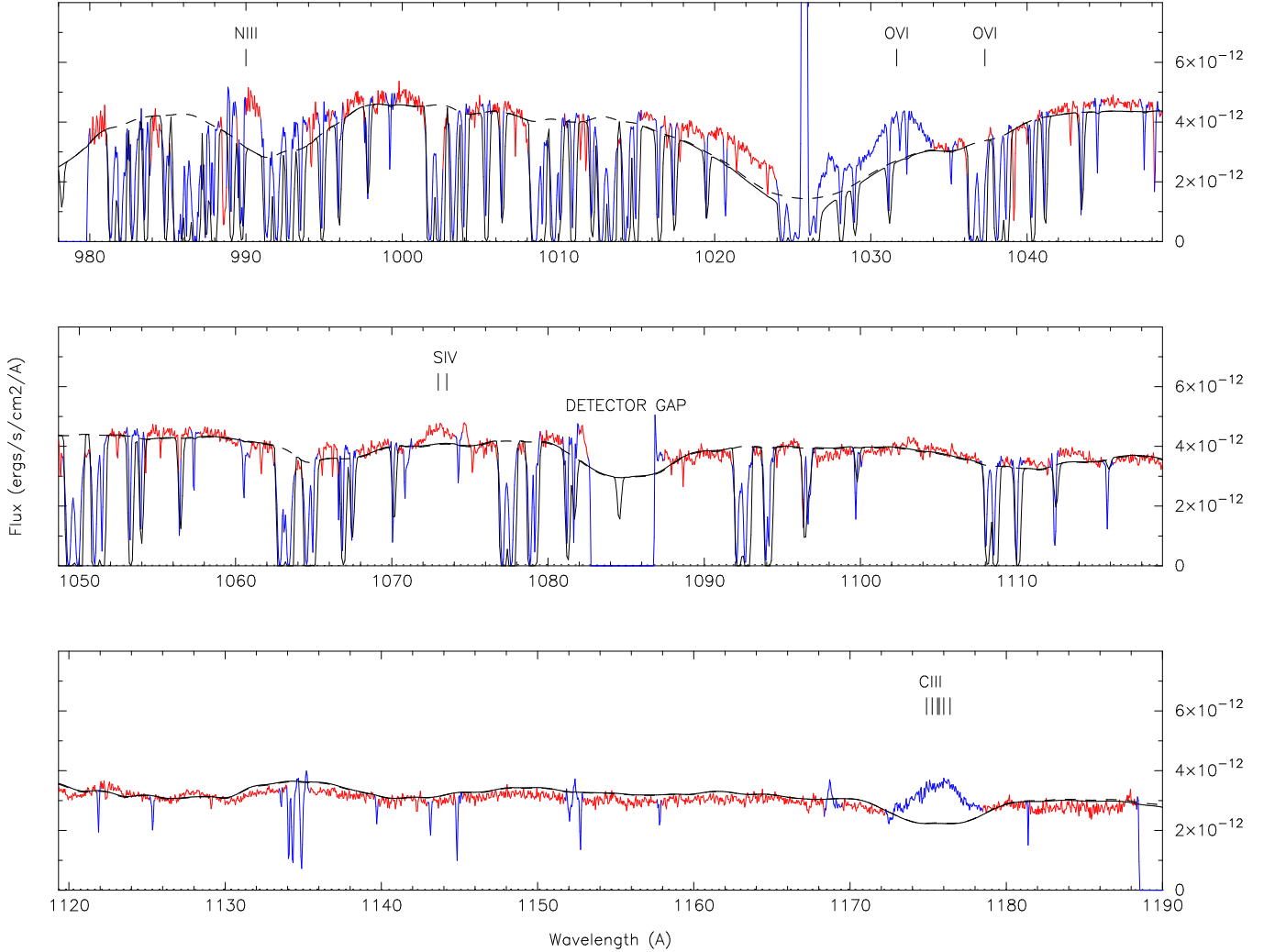} 
\caption{ 
Best fitting disk model to the FUSE spectrum of V603 Aql 
assuming $M_{wd}=1.0M_{\odot}$, d=250pc, and $i=18^{\circ}$. 
The observed FUSE spectrum is in red and blue, the model 
is represented with the solid black line. Fitting is performed 
between the red line and the solid black line, the portions of
the spectrum masked before the fitting are shown in blue.   
The model consists of an accretion disk with $\dot{M}=2.4 \times 10^{-9}
M_{\odot}$/yr. All the sharp absorption lines, believed to be due 
to the ISM, have been masked out and are colored in blue.  While the
obvious O\,{\sc vi} (1031.9\AA\ ) and C\,{\sc iii} (1075\AA\ ) lines
were mask before the fitting, the non-obvious N\,{\sc iii} (990\AA\ ) 
and S\,{\sc iv} (1073\AA\ ) lines were not masked and became apparent
only in the fitting. The modeling includes an elementary ISM model
which reproduces the ISM absorption.   The model is also shown without
the ISM absorption with the dashed black line.  The inclusion of a 
WD model to the fitting did not produce a significant change to the
fitting as it contributed to only $\sim$1\% of the flux.  
} 
\end{figure} 

\begin{figure} 
\vspace{-5.cm} 
\plotone{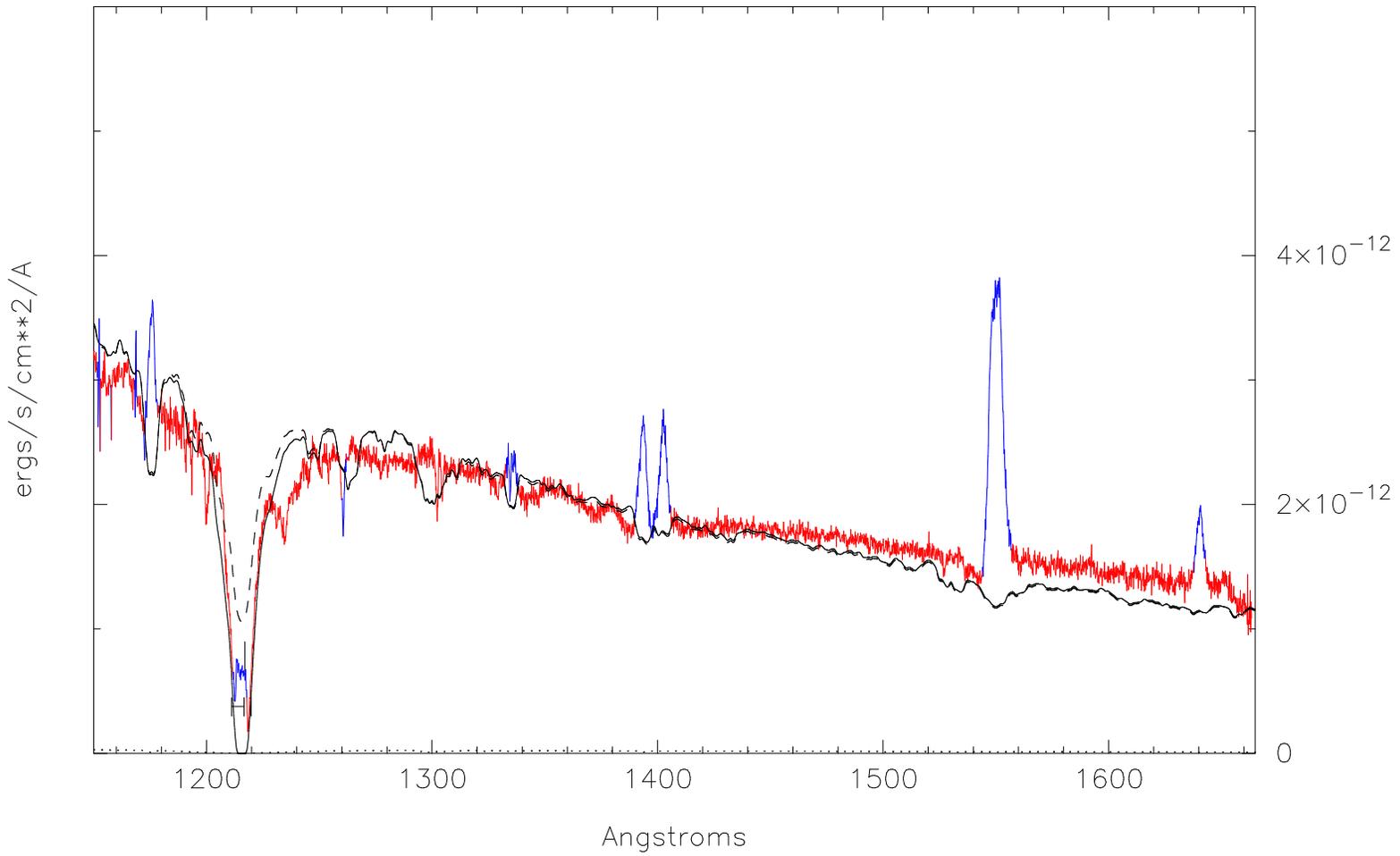}  
\caption{ 
Best fitting disk model to the HST/GHRS spectrum of V603 Aql. 
This model is for a WD with a mass $M_{wd}= 1 M_{\odot}$, it has 
an inclination $i=18$deg, mass accretion rate $\dot{M}=3.0 \times 
10^{-9}M_{\odot}$/yr.  The distance is 250pc. 
The dashed line shows the Ly$\alpha$ region 
{\it without} the ISM modeling. The regions
that have been masked (and that are not modeled) for the fitting 
are in blue.  
} 
\end{figure} 

\begin{figure}
\vspace{-5.cm} 
\plotone{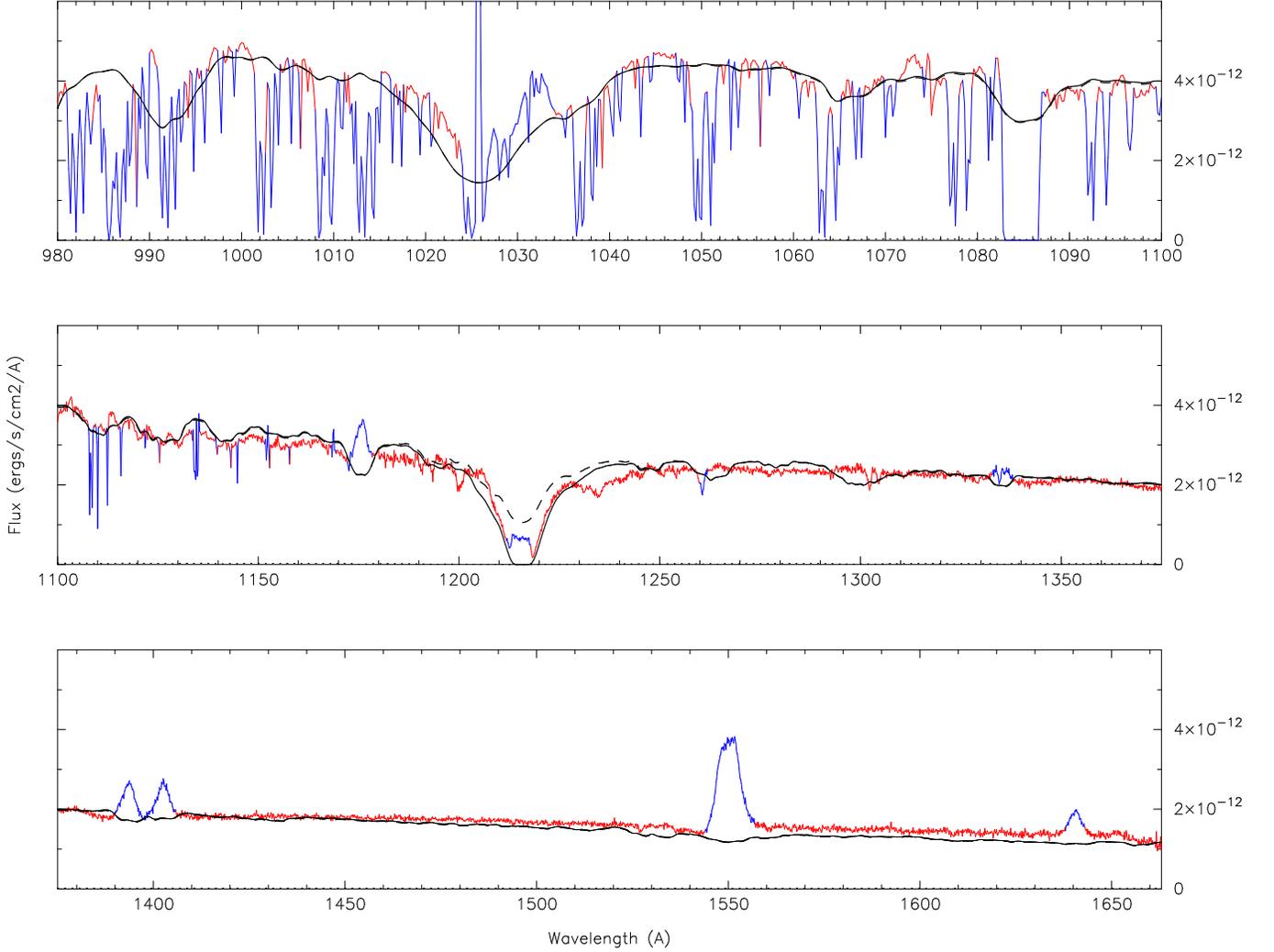}  
\caption{ 
Best Fitting disk model to the FUSE + GHRS wavelength range of V603 Aql. 
This model is for a WD with a mass $M_{wd}= 1 M_{\odot}$, it has 
an inclination $i=18$deg, mass accretion rate $\dot{M}=2.6 \times 
10^{-9}M_{\odot}$/yr. The GHRS spectrum has been scaled to the
FUSE spectrum using a factor of 0.80. The distance is 250pc. 
For clarity, the ISM model has been omitted from the FUSE spectral
range but it has been kept for the Ly$\alpha$ region. The dashed line
shows that region {\it without} the ISM modeling. The regions
that have been masked (and that are not modeled) for the fitting 
are in blue.  
} 
\end{figure}


\begin{thebibliography}{} 

\bibitem[Arenas et al. (2000)]{are00} 
Arenas et al. 2000, \mnras, 311, 135


\bibitem[Borczyk et al.(2003)]{bor03} 
Borczyk, W., Schwarzenberg-Czerny, A., \& Szkody, P. 2003, \aap, 405, 663  

\bibitem[Duerbeck (1987)]{due87} 
Duerbeck, H. 1987, \apss, 131, 461

\bibitem[Fitzpatrick (1999)]{fit99}
Fitzpatrick, E.L., 1999, PASP, 11, 63

\bibitem[Friedjung et al. (1997)]{fri97}
Friedjung, M., Selvelli, P., Cassatella, A.  1997, A\&A, 318, 204 

\bibitem[Gallagher \& Holm (1974)]{gal74} 
Gallagher, J. S. \& Holm, A. V. 1974, \apj, 189, L123

\bibitem[Godon et al. (2012)]{god12} 
Godon, P., et al. 2012, \apjs, 203, 29 


\bibitem[Harrison \& Gehrz (1988)]{har88} 
Harrison \& Gehrz 1988, \aj, 96, 1001 

\bibitem[Harrison et al. (2013)]{har13} 
Harrison et al. 2013, \apj, 767, 7

\bibitem[Hubble \& Duncan (1927)]{hub27} 
Hubble, E. \& Duncan, J.C., 1927, \apj, 66, 59

\bibitem[Hubeny (1988)]{hub88} 
Hubeny, I. 1988, Computer Physics Communications, 52, 103 

\bibitem[Hubeny \& Lanz (1995)]{hub95} 
Hubeny, I., \& Lanz, T. 1995, \apj, 439, 875 

\bibitem[Johnson et al.(2014)]{joh14}
Johnson, C.B., Schaefer, B.E., Kroll, P., Henden, A.A. 2014, ApJ, 780, 25 

\bibitem[McLaughlin (1960)]{mcl60} 
McLaughlin, D.B. 1960, in Stellar atmospheres, Ed. by J.L. 
Greenstein,(Chicago: University of Chicago Press), p.585 

\bibitem[Mukai \& Orio (2005)]{muk05} 
Mukai, K. \& Orio, M. 2005, \apj, 622, 602

\bibitem[Puebla et al. (2007)]{pue07} 
Puebla, R.E., Diaz, M.P., Hubeny, I. 2007, \aj, 134, 1923

\bibitem[Popham \& Narayan (1995)]{pop95} 
Popham, R., \& Narayan, R. 1995, \apj, 442, 337 


\bibitem[Ritter \& Kolb (1998)]{rit98} 
Ritter, H., \& Kolb, U. 1998, \aap, 129, 83


\bibitem[Savage \& Mathis (1979)]{sav79}
Savage, B.D., \& Mathis, J.S.  1979, ARA\&A, 17, 73

\bibitem[Schaefer (2014a)]{sch14a}
Schaefer, B. 2014a, BAAS,224, 412.02

\bibitem[Schaefer (2014b)]{sch14b}
Schaefer, B. 2014b, private communication, Columbia
CV Workshop, December, 2014 

\bibitem[Seaton (1979)]{sea79}
Seaton, M.J., 1979, MNRAS,  187, 73

\bibitem[Selvelli et al. (1989)]{sel89} 
Selvelli et al. 1989 Lecture Notes in Physics, vol.369, IAU Colloq.122      

\bibitem[Selvelli \& Gilmozzi (2013)]{sel13} 
Selvelli, P., \& Gilmozzi, R. 2013, \aap, 560, 49



\bibitem[Shara et al. (1986)]{sha86} 
Shara, M., Livio, M., Moffat, A., \&  Orio, M. 1986, \apj, 311, 163 

\bibitem[Tappert et al. (2013)]{tap13} 
Tappert et al. 2013, \mnras, 431, 92


\bibitem[Verbunt (1987)]{ver87} 
Verbunt, F. 1987, \aaps, 71, 339 

\bibitem[Wade \& Hubeny (1998)]{wad98} 
Wade, R., \& Hubeny, I. 1998, \apj, 509, 350

\end{thebibliography}
\end{document}